\documentclass[twocolumn,aps,showkeys,amsmath,amssymb]{revtex4}

\usepackage{graphicx}
\usepackage{dcolumn}
\usepackage{bm}

\begin{document}

\title{Pressure correction in density-functional calculations}

\author{Shun Hang Lee}\email{shlee@phy.cuhk.edu.hk}

\affiliation{Department of Physics, The Chinese University of Hong Kong,
Shatin, New Territories, Hong Kong.}

\author{Jones Tsz Kai Wan}\email{jwan@phy.cuhk.edu.hk}

\affiliation{Department of Physics, The Chinese University of Hong Kong,
Shatin, New Territories, Hong Kong.}

\date{\today}

\begin{abstract}
First-principles calculations based on density functional theory have been 
widely used in studies of the structural, thermoelastic, rheological, and 
electronic properties of earth-forming materials. The exchange-correlation 
term, however, is implemented based on various approximations, and this is 
believed to be the main reason for discrepancies between experiments and 
theoretical predictions.
In this work, by using periclase MgO as a prototype system we examine the 
discrepancies in pressure and Kohn-Sham energy that are due to the choice of the
exchange-correlation functional. For instance, we choose local density 
approximation and generalized gradient approximation. We perform extensive 
first-principles calculations at various temperatures and volumes and 
find that the exchange-correlation-based discrepancies in Kohn-Sham energy and 
pressure should be independent of temperature. This implies that the physical 
quantities, such as the equation of states, heat capacity, and the 
Gr\"{u}neisen parameter,
estimated by a particular choice of exchange-correlation functional can easily 
be transformed into those estimated by another exchange-correlation functional. 
Our 
findings may be helpful in providing useful constraints on mineral properties 
at deep Earth thermodynamic conditions.
\end{abstract}

\keywords{pressure correction, density functional theory, exchange-correlation, Car-Parrinello molecular dynamics.}

\maketitle

\newcommand{\supers}[1]{$^\textrm{#1}$}
\newcommand{\subs}[1]{$_\textrm{#1}$}
\newcommand{\supersub}[2]{$^\textrm{#1}_\textrm{#2}$}

\section{Introduction}
In recent decades, first-principles (FP) calculations with density functional 
theory (DFT) \cite{hohenberg_1964,kohn_1965} have been widely adopted in studies of the structural, 
thermoelastic, rheological, and electronic properties of materials, 
molecules, and minerals in research areas that range from physics and 
chemistry to biology and geosciences
\cite{allen_1993,wentzcovitch_prl_1993,parrinello_1980,car_qsar_2002,carloni_qsar_2002,oganov_zkrist_2005}.
The modern implementation of FP 
techniques is based on different approximations, such as the pseudopotential
(PP) method, the expansion of electronic wavefunctions by modal functions, and 
the choice 
of exchange-correlation (XC) functionals. As a result, the predictions made by 
different FP calculations may differ from one another moderately.

Many techniques have been developed to enhance the accuracy and efficiency of 
FP calculations. Examples include the projector-augmented wave (PAW) method 
\cite{blochl_prb_1994}, and maximally localized Wannier functions
\cite{mostofi_cpc_2008}. However, due to inadequate knowledge, the exact form 
of exchange-correlation functional remains unknown and the search for a 
precise XC term has been an active area of research. The choice of XC in FP 
calculations has been shown to be the main reason for the inconsistencies 
between these various calculations
\cite{karki_am_1997,karki_prb_2000,oganov_epsl_2001,oganov_nature_2001,stixrude_science_2005,wan_jgr_2007}.
In particular, the pressures calculated by generalized gradient 
approximation (GGA) \cite{perdew_1996} and local density approximation (LDA) \cite{ceperley_prl_1980} are known to be 
systematically over- and underestimated, and a common way to choose the XC 
functional is to use the XC that gives the best agreement with experiments.

In this article, we investigate the way in which such a pressure difference 
depends on 
volume and temperature. If this difference can be quantified, simulations can 
be performed using either exchange-correlation, and at the same time, obtaining 
the results of the other. This would be an efficient way of constraining the 
thermodynamic quantities. For example,
\citet{karki_am_1997} 
calculated the elastic properties of $\text{MgSiO}_3$ at pressures up to the 
lower mantle 
condition using LDA, which 
gave an equilibrium volume and elastic constant consistent with experiments
\cite{yeganeh_pepi_1994,Ross_pcminer_1989}. However, calculations by
\citet{oganov_epsl_2001,oganov_nature_2001}
on $\text{MgSiO}_3$
have shown 
that the GGA gives an extremely accurate elastic constant and the volume 
dependencies of the elastic properties, although the pressure is 
overestimated. The authors thus applied a constant pressure shift to the 
equation of state (EOS) estimated by GGA and then matched the measured ambient 
pressure EOS, which resulted in an excellent match between the experimental 
observations and the FP calculation results.

\section{theory}
To quantify the relationship between the pressure difference and the XC 
functionals, we first consider the total internal energy of a system, which is 
contributed by the interaction energy between particles and the total kinetic 
energy of the ions and electrons. At thermal equilibrium, this total internal 
energy is given by
\begin{eqnarray}
U &=&{\Big \langle} \sum_I \frac{1}{2}M_I \dot{R}_I^2 {\Big \rangle}
+{\Big \langle} \sum_{I,J\neq I}\frac{1}{2} \frac{q_I q_J}{|{\bf R}_I - {\bf R}_J|}{\Big \rangle}\nonumber \\
&+&{\Big \langle} E_{\rm KS}[\{ {\bf R}_I\},\{ \psi_n\}]{\Big  \rangle},\label{total.E}
\label{total energy}
\end{eqnarray}
where $\langle\cdots\rangle$ denotes the ensemble average, and $\{ \mathbf{R}_I \}$ 
and $\{ \psi_n \}$ represent ionic and electronic 
degrees of freedom. Also, $M_I$ and $q_I$ represent the mass and charge of the 
$I$th 
ion. The last term $E_{\rm KS}[\lbrace \mathbf{R}_I \rbrace ,\lbrace \psi_n \rbrace]$ 
is the Kohn-Sham energy functional of the 
ion-electron system. Within adiabatic approximation, 
$E_{\rm KS}[\lbrace \mathbf{R}_I \rbrace ,\lbrace \psi_n \rbrace]$ is given by 
DFT, and $E_{\rm KS}$ depends on 
the configuration of ions for bulk systems.
The first and the second terms in 
Eq.~(\ref{total energy}) are the classical kinetic and coulomb energy of the 
ions.
For a crystal system with small oscillation of ions,
one can write ${\bf R}_I = {\bf R}^0_I+\delta{\bf R}_I$,
where ${\bf R}^0_I$ and $\delta{\bf R}_I$ are respectively
the equilibrium position and the (small) displacement from ${\bf R}^0_I$
of the $I$th ion.
In addition, we can assume that each ion is under the influence of an
effective local potential,
given by $(1/2)K_{I,\alpha}(\delta R_{I,\alpha})^2$.
Here $K_{I,\alpha}$ is the effective force constant,
which includes the contributions from ion-ion and electron-ion interactions,
and $\alpha$ is the component of displacement ($\alpha = 1,2,3$).
As a result,
the total internal energy [Eq.~(\ref{total energy})] can be approximated by
\begin{eqnarray}
U &\approx&{\Big \langle} \sum_I \frac{1}{2}M_I \delta\dot{R}_I^2 {\Big \rangle}
+\sum_{I,J\neq I}\frac{1}{2} \frac{q_I q_J}{|{\bf R}^0_I - {\bf R}^0_J|}\nonumber \\
&+&E_{\rm KS}[\{ {\bf R}^0_I\},\{ \psi^0_n\}]
+{\Big \langle}\sum_{I,\alpha}\frac{1}{2}K_{I,\alpha}(\delta R_{I,\alpha})^2{\Big \rangle}\nonumber \\
&=& \langle T \rangle + U^0_{\rm ion} + E^0_{\rm KS} + \langle \delta E\rangle,
\label{total.E.approx}
\end{eqnarray}
which could be understood as the sum of
zero-temperature energies (quantities with superscript 0) and
the contribution due to thermal motions of ions (quantities with
$\langle\cdots\rangle$).

In first-principles calculations, $E^0_{\rm KS}$ and $\langle \delta E \rangle$
depend on the choice of XC functional, 
whereas $\langle T \rangle$ equals $3Nk_BT/2$ regardless of the choice of XC,
because the ionic motions are treated classically.
Therefore, Eq.~(\ref{total.E.approx}) can be written as,
\begin{equation}
U = \frac{3}{2}Nk_BT + \langle E\rangle,
\end{equation}
where
\begin{equation}
\langle E\rangle =  U^0_{\rm ion} + E^0_{\rm KS} + \langle \delta E\rangle,
\end{equation}
is the total DFT energy of the system.
We now consider the internal energy calculated by two different choices of XC,
say XC$_1$ and XC$_2$, the difference between $U$ is given by
\begin{eqnarray}
\Delta U(V,T)
&=&U_{{\rm XC}_1}(V,T)-U_{{\rm XC}_2}(V,T)\nonumber \\
&=&\langle E_{{\rm XC}_1}-E_{{\rm XC}_2}\rangle.
\label{delta.E.VT}
\end{eqnarray}
As $U^0_{\rm ion}$ only depends on the equilibrium ionic configuration, it
does not contribute to $\Delta U(V,T)$. Moreover,
$\langle \delta E\rangle$ should be equal to $3Nk_BT/2$, because $\delta R_{I,\alpha}$
is regarded as a classical degree of freedom and
$$
\langle\frac{1}{2}K_{I,\alpha}(\delta R_{I,\alpha})^2 \rangle= \frac{1}{2}k_BT,
$$
at thermal equilibrium. Therefore, $\Delta U$ becomes
\begin{eqnarray}
\Delta U(V,T)
&=& E^0_{{\rm KS},{\rm XC}_1}-E^0_{{\rm KS},{\rm XC}_2},
\label{delta.E.V}
\end{eqnarray}
which depends solely on the system volume.
More importantly, the difference between the calculated pressures, given by
\begin{equation}
P_{{\rm XC}_1}(V,T)-P_{{\rm XC}_2}(V,T)
=-\frac{\partial \langle E_{{\rm XC}_1}-E_{{\rm XC}_2}\rangle}{\partial V}{\bigg |}_S,
\label{delta.p.VT}
\end{equation}
(where $S$ represents the entropy of the system)
should be, in general, a function of $V$ and $T$. However, according to 
Eqs.~(\ref{delta.E.VT}) and (\ref{delta.E.V}), 
although both $E_{{\rm XC}_1}$ and $E_{{\rm XC}_2}$ depend on temperature, 
their difference, 
$E_{{\rm XC}_1}-E_{{\rm XC}_2}$ should not. Consequently,
the resultant difference in pressure should also be
temperature independent, that is,
\begin{equation}
P_{{\rm XC}_1}(V,T)-P_{{\rm XC}_2}(V,T)=\Delta P_{12}(V).
\label{delta.p.V}
\end{equation}

\section{Methodology}
To illustrate the relationships in Eqs.~(\ref{delta.E.V}) and (\ref{delta.p.V}),
we perform FP calculations on periclase 
MgO at thermodynamic conditions that range from ambient conditions to pressures 
and temperatures that are close to those of a deep Earth environment. Periclase MgO has a 
face-centered cubic structure and is known to be stable in the pressures and 
temperatures being considered in this work. Therefore, structural changes, such 
as phase transition and melting, are avoided.

The EOS at each temperature are determined by 
first-principles molecular dynamics simulations. For each simulation, the 
simulation supercell consists of 64 atoms (32 MgO units). These atoms were 
first 
arranged in a face-centered cubic (fcc) configuration, and the atomic 
trajectories and 
electronic orbitals were evolved via Car-Parrinello molecular dynamics (CPMD) 
\cite{car_1985}, with the temperature controlled by the Nos\'{e} thermostat technique
\cite{nose_1984}, at $T$ = 300~K, 1000~K, 2000~K,... 4000~K, at various cell 
volumes. The calculations are repeated using both LDA and GGA 
exchange-correlations. The pressures determined at each temperature are fitted
against the third-order Birch-Murnaghan EOS \cite{birch_pr_1947}.
As a result, two EOSs at each temperature are obtained, one LDA and one GGA.

The choice of the XC energy functional is implemented in the 
generation of the pseudopotentials. In this study, two pairs of 
pseudopotentials for Mg and O are generated, one pair generated using LDA 
and the other using GGA. Apart from this difference in XC, all of the other 
parameters, such as the cutoff radius and the valence states, are the 
same. For Mg, a norm conserving pseudopotential \cite{troullier_1991} with 
nonlinear core correction \cite{louie_1982} is used, whereas for O an 
ultrasoft \cite{vanderbilt_1990} pseudopotential is used. The reference 
configuration is 3s1.5 3p0 3d0 for Mg and 2s2 2p2 for O. For CPMD, a 
fictitious electron mass $\mu_e=400$ $m_e$ and a simulation time step of 
$\Delta t = 12$~
atomic time units ($\sim0.3$ fs) are used.
Electronic wavefunctions are expanded by planewaves with a cutoff of 30 Ry,
and the cutoff for charge density is 240 Ry.
To ensure enough statistical data, 
the simulations are run for more than 4~ps at each $V$, $T$ point.

\section{Results and discussion}
\subsection{Equation of states and bulk modulus}
The EOS given by our LDA and GGA calculations are shown in 
Fig.~1. The results are fitted against the third-order Birch-Murnaghan EOS, with 
$V_0 = 3782.30$ $a_B^3$, $K_0 = 177.486$~GPa, and $K_0' = 4.026$ for LDA and 
$V_0 = 4046.43$ $a_B^3$, $K_0 = 149.320$~GPa, and $K_0' = 4.080$  for GGA.
To ensure that our calculations are 
compatible with existing results, we compare our results with those obtained by 
\citet{karki_prb_mgo_2000},
and
\citet{isaak_jgr_1990}.
The bulk moduli for each EOS are shown in Fig.~2.
We recalculate the static EOS of
\citet{karki_prb_mgo_2000}
by using the same PP files as in their work \cite{karki_pp_pwscf},
and the recalculated static EOS is in excellent agreement with their published EOS parameters.
In addition, such EOS was shown to be highly consistent with the
experimental measurements at moderate pressures ($P < 170$ GPa). Therefore, 
Fig.~1 and Fig.~2 illustrate the discrepancy that arises from different 
methodologies, such as the implementation of PP, the choice of XC functionals,
and the interaction potentials.
\begin{figure}[htb]
\includegraphics[width=0.9\columnwidth]{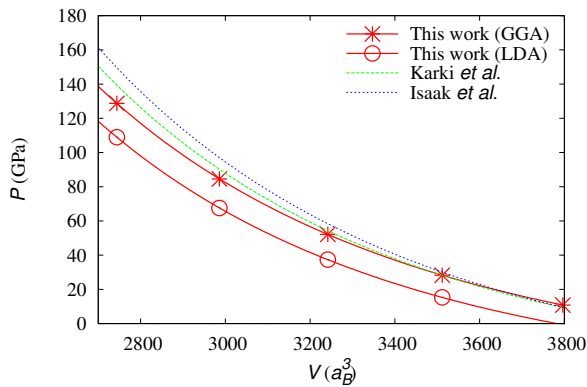}
\caption{(Color online) Calculated equations of state for MgO by LDA (circle) and GGA (star) at 0 K.
The results of
\citet{karki_prb_mgo_2000}
(dashed line) 
and
\citet{isaak_jgr_1990}
(dotted line) are given for comparison.
The EOSs are drawn using the published EOS parameters.}
\label{fig.1}
\end{figure}

\begin{figure}[htb]
\includegraphics[width=0.9\columnwidth]{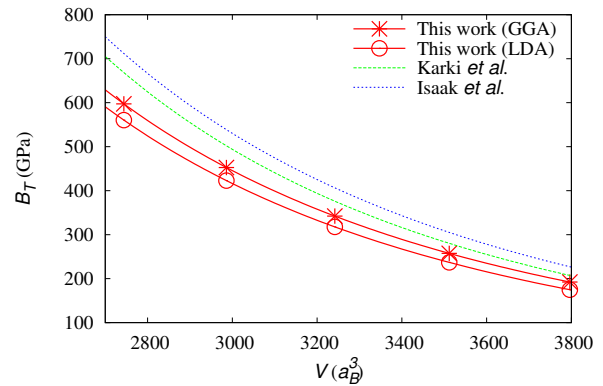}
\caption{(Color online) Isothermal bulk moduli given by various calculations.}
\label{fig.2}
\end{figure}

\begin{figure}[htb]
\includegraphics[width=0.9\columnwidth]{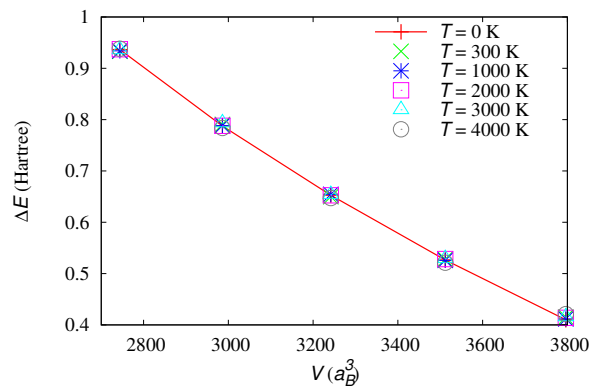}
\caption{(Color online) DFT total energy difference between GGA and LDA against volume at different temperatures.
The results for $T=0$ K are joined for visualization.}
\label{fig.3}
\end{figure}

\begin{figure}[htb]
\includegraphics[width=0.9\columnwidth]{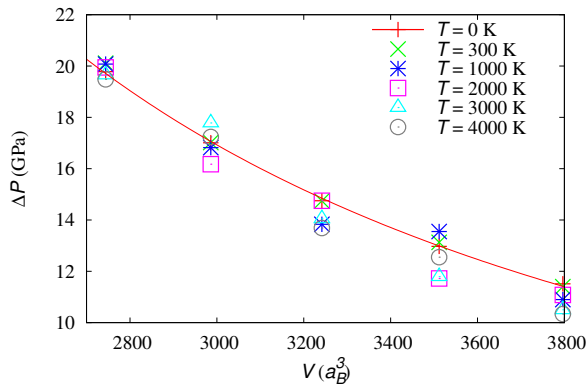}
\caption{(Color online) Difference in pressure $\Delta P = P_{\rm GGA}-P_{\rm LDA}$ at each 
temperature. Note that the pressure difference is in the order of 10 GPa, and 
the maximum difference between these points is about 1 GPa, which is negligible 
in DFT-based calculations. The pressure difference for $T=0$ K is fitted against
$\Delta P=a_1/V+a_2/V^2$ with
$a_1=1.53\times 10^4$ GPa $a_B^3$ and
$a_2=1.06\times 10^8$ GPa $a_B^6$ for visualization.}
\label{fig.4}
\end{figure}

At first glance, this discrepancy appears to reduce as the system volume 
increases. This is because both the pseudopotential errors and the difference 
between the 
XC functionals become less significant as atomic separation increases. Also, 
the GGA-calculated pressure and the bulk modulus in this work deviate 
systematically from those of LDA. In addition, our GGA results are more 
compatible with those estimated in
Ref.~27
and
Ref.~28
than our 
LDA calculations. This indicates that the GGA calculation should, 
in principle, be closer to the experimental measurements.
The Mg PP used in Ref.~27
was optimized by using a combination of various electronic configurations,
and a norm-conserving O PP was used.
Nevertheless, the optimization of a PP 
configuration is the subject of further study. We concentrate on the errors 
introduced by the XC alone.

\subsection{Energy and pressure differences}
According to Eq.~(\ref{delta.E.VT}), 
it is intuitive to calculate the difference in 
energy at each temperature and volume. In Fig.~3, we plot the difference in 
energy, 
$\Delta E = E_{\rm GGA} - E_{\rm LDA}$,
against $V$ at each temperature and find that it 
does not depend on temperature. This is consistent with 
the prediction of Eq.~(\ref{delta.E.V}). 
The differences between the pressures estimated by LDA 
and GGA at each temperature, 
$\Delta P = P_{\rm GGA} - P_{\rm LDA}$,
are shown in Fig.~4. As can be seen, the 
differences in pressure at all temperatures almost coincide, with a maximum 
deviation of about 1~GPa, which is much smaller than the typical statistical 
error 
in pressure estimated by the DFT-based calculations. This is again in 
accordance with the conjecture of Eq.~(\ref{delta.p.V}), and the pressure
difference should be independent of temperature.
Within the pressure range we have examined (0 GPa $<P<$ 170 GPa),
$\Delta P(V)$ behaves asymptotically as a polynomial of $1/V$, that is,
\begin{equation}
\Delta P(V)\simeq \sum_{n=1}^{\infty}a_n {\left( \frac{1}{V} \right)}^n.
\label{series}
\end{equation}

Here, a few comments are in order. The present analysis implies that the
thermodynamic quantities, such as internal energy and pressure, calculated by FP 
methodologies with a particular XC functional can easily be transformed into 
those obtained by another choice of XC functional. For example, the 
difference in heat capacity, given by
\begin{eqnarray}
\Delta C_V &=& C_{V,{\rm GGA}} - C_{V,{\rm LDA}} \nonumber\\
&=& 
\frac{\partial \langle E_{\rm GGA} - E_{\rm LDA} \rangle }{\partial T} = 
\frac{\partial \langle \Delta E \rangle}{\partial T} = 0,
\end{eqnarray}
at each volume should vanish because $\Delta E$ does not depend on 
temperature. This implies that the heat capacity calculated by any XC should 
be the same.
In addition, the Gr\"{u}neisen parameter,
\begin{equation}
\gamma = \frac{\alpha K_T V}{C_V},
\label{gruneisen}
\end{equation}
calculated by different XC functionals should also be the same.
Here $\alpha=(1/V)(\partial V/\partial T)_P$ and $K_T=-V(\partial P/\partial V)_T$
are, respectively, the coefficient of thermal expansion and the isothermal bulk modulus.
To prove the statement, we consider the following thermodynamic relation,
\begin{equation}
\bigg(\frac{\partial P}{\partial T}\bigg)_V
= -\bigg(\frac{\partial V}{\partial T}\bigg)_P
\bigg(\frac{\partial P}{\partial V}\bigg)_T = \alpha K_T.
\end{equation}
Therefore, Eq. (\ref{gruneisen}) can be written as
\begin{equation}
\gamma = \frac{V}{C_V}\frac{\partial P}{\partial T}.
\end{equation}
As a result, the difference in the Gr\"{u}neisen parameter,
\begin{eqnarray}
\Delta\gamma(V,T) &=&
  \frac{V}{C_{V,{\rm GGA}}(V,T)}\frac{\partial P_{\rm GGA}(V,T)}{\partial T} \nonumber\\
  &-&\frac{V}{C_{V,{\rm LDA}}(V,T)}\frac{\partial P_{\rm LDA}(V,T)}{\partial T} \nonumber\\
&=& \frac{V}{C_V(V,T)}\frac{\partial\Delta P(V,T)}{\partial T} = 0,
\label{gamma}
\end{eqnarray}
should vanish either.

The present estimation of pressure and energy differences is based on 
calculations of the cubic MgO structure. In principle, our analysis should be 
valid 
for any crystal system with small oscillation of ions,
although the corresponding FP calculations may require 
extra 
care in the estimation of pressure.
Although not shown in this work, our preliminary calculations on MgSiO$_3$
perovskite and post-perovskite phases \cite{lee.mgsio3.private} support our proposed theory.
For a non-cubic crystal structure, a finite 
strain may introduce shear stress to the system. In such a case, the EOS should 
be determined by cell dynamics algorithms
\cite{wentzcovitch_prl_1993,parrinello_1980}.
It should also be noted that our conclusion on the pressure correction
relies on the assumption that the thermal contribution to the total
internal energy is governed by classical mechanics. However, this may not
be valid in situations where quantum mechanical interactions, such as
phonon vibration should be taken into account. In particular, at extremely
low temperatures, long range acoustic phonon modes play an important role
in various thermodynamic phenomena.
Nevertheless, recent works on structural phase transition in MgSiO$_3$
at core-mantle boundary conditions \citep{tsuchiya_epsl_2004}
and postspinel transition in Mg$_2$SiO$_4$ \citep{yu.grl.mg2sio4.2007}
have shown that high-temperature thermodynamic properties
estimated by phonon-based calculations should not depend on the choice
of XC functionals, as long as quasiharmonic approximation (QHA) remains valid
in the thermodynamic conditions of interest.
For example, different choices of XC functionals only affect the position of the
estimated phase boundaries in these works but not the Clapeyron slopes.
In addition, the recent theoretical study of ultrahigh-pressure EOS of
MgO \citep{wu.jgr.mgo.2008} has also shown explicitly that,
within the QHA validity regime, the difference between
a LDA-EOS and a GGA-EOS is independent of temperature.
More importantly, it should be noted that QHA requires the vibrational amplitude of
each atom to be small, which is the major assumption of our pressure correction conjecture.
As a result, the work of \citet{wu.jgr.mgo.2008} ubiquitously supports our hypothesis of
pressure correction, even the FP methodologies used (lattice dynamics) in this work
are different from ours (Car-Parrinello molecular dynamics).

The validity and behavior of $\Delta P(V)$ are also subjects for further 
investigation. In the pressure range we have investigated, $\Delta P(V)$ is 
finite and is well approximated by Eq.~(\ref{series}), which implies that the 
pressure 
difference should increase with a decreasing volume. However, this is not valid 
either when $V$ approaches infinity ($V\rightarrow\infty$) or when $V$ becomes 
vanishingly small ($V\approx 0$). In the former case, the system consists of 
isolated atoms, and the energy difference that is due to different XC is 
constant; as a 
result $\Delta P(V)$ is zero as $V\rightarrow\infty$. In the latter case, the 
inter-atomic distance approaches zero, the interacting electrons should be well 
described by the free electron gas model, and LDA and GGA should give the same 
pressure.
Last but not the least, the present study focuses on the pressure correction
due to the chosen XC functional, which is in opposite to the idea of 
{\it volume correction} suggested by
\citet{wu.jgr.mgo.2008},
in which an EOS is corrected to match experimental data.
A combination of these EOS corrections will make various FP-based
results more transferable, thus allowing one to have an effective recipe to transform
the calculated thermodynamic quantities from one condition to another. 
\section{Conclusion}
To conclude, by using MgO as a prototype system, we have studied the  
discrepancy in pressure that is due to different choices of XC functionals in 
DFT calculations. We have found that the differences in energy and 
pressure for GGA and LDA calculations should be independent of temperature. As 
a result, one may easily constrain the XC error at arbitrary 
temperatures. This may lead to a better estimation of thermodynamic properties, 
such as heat capacity and the Gr\"{u}neisen parameter, in the systems that are 
being investigated by the 
FP and high-pressure communities.

\section*{ACKNOWLEDGEMENTS}
The authors acknowledge help from S. C. Sung and a useful discussion with A. R. 
Oganov. S. H. Lee acknowledges the support of S. M. Wong and S. F. Tsang.
Jones T. K. Wan acknowledges the support of S. S. Lam and T. L. 
Wan. First-principles calculations were performed using
the Car-Parrinello molecular dynamics code in Quantum-ESPRESSO ver. 
3.1.1. Computation was performed using the CUHK high-performance computing 
(HPC) facility. This work is supported by RGC-HK (CERG project no. 403007) and 
CUHK (direct grant no. 2060307) grants.



\end{document}